\documentclass[12pt]{article}
\usepackage{amsfonts,amsmath}
\usepackage[mathscr]{eucal}
\usepackage{amssymb}
\usepackage{amsthm}
\theoremstyle{plain}

\usepackage{epsfig,epsf}
\usepackage{graphicx, subfigure}
\usepackage{graphicx}

\newcommand{\be}{\begin{eqnarray}}
\newcommand{\ee}{\end{eqnarray}}
\newcommand{\nn}{\nonumber \\}
\newcommand{\lb}{\label}
\newcommand{\p}[1]{(\ref{#1})}

\begin{document}

\begin{titlepage}

\vfill
\vfill

\begin{center}
\baselineskip=16pt {\Large  Supersymmetric field theory with benign ghosts}
\vskip 0.3cm {\large {\sl }}
\vskip 10.mm {\bf 
A.~V. Smilga}
 \\
\vskip 1cm

SUBATECH, Universit\'e de Nantes, \\
4 rue Alfred Kastler, BP 20722, Nantes 44307, France
\footnote{On leave of absence from ITEP, Moscow, Russia}\\
E-mail:  smilga@subatech.in2p3.fr

\end{center}

\vspace{0.2cm} \vskip 0.6truecm \nopagebreak

\begin{abstract}
\noindent
We construct a supersymmetric 1+1 - dimensional field theory involving extra derivatives and associated
ghosts: the spectrum of the Hamiltonian is not bounded from below, neither from above.
In spite of that, there is neither classical, nor quantum collapse and unitarity is preserved. 

\end{abstract}


\end{titlepage}

\section{Introduction}

We still do not know what quantum gravity is. One of the problems
\footnote{Classical and quantum  gravity suffer also from other problems associated with the 
geometric nature of the theory and the absence of the flat universal time. These problems 
(first of all the problem of causality violation)  are not less troubling \cite{Isham}. We address the reader
to Ref. \cite{dragon} for a detailed discussion of these and related issues.}
is that the quantum version of Einstein gravity is not renormalizable: the expansion runs over $G\Lambda^2$
($G$ being the Newton constant and $\Lambda$  the ultraviolet cutoff) and  severe power divergences are manifest.
 This is probably also true for Einstein supergravity
\footnote{At the moment, one can still hope that the maximal 
${\cal N} =8$ extended supergravity is finite.
Remarkable cancellations through at least four loops  were observed in 
\cite{Bern}. These cancellations are not without
a reason \cite{Bossard}. However, most theorists expect that the divergencies 
would pop up at the 7-th loop 
level and higher \cite{Beisert,Cederwall}.}

On the other hand,  the higher-derivative gravity   involving the structures $R^2$ and $R_{\mu\nu}^2$ 
in the Lagrangian
has  dimensionless coupling constants and is renormalizable \cite{Stelle}. A supersymmetric version of this 
theory is even asymptotically free \cite{Fradkin}. One can imagine then a scenario where the 
fundamental gravity theory involves higher derivatives, with the usual gravity of our World appearing
 as an effective
low-energy theory \cite{Sakharov,Zvenigorod,Adler}. 

Recently, there was a revival of interest to higher derivative gravity (see e.g. \cite{Nieder,Mann,Anca}). 
Still, 
this scenario has  mostly not been considered attractive by scientific community because 
higher-derivative theories have been known since \cite{PU} to be ghost-ridden. 
Ghosts are usually conceived as the poles with 
the wrong residue in propagators. The associated states have negative norm  and the common lore is that 
their production violates unitarity.

 It is important to understand, however, that one {\it is} able to quantize theory such that the norm of all states 
remains positive. 
The price for that is that the states with arbitrary negative energy exist such that the Hamiltonian
does not have a ground state. The perturbative vacuum is absent. 
At the level of {\it free} Hamiltonian, this is not  a problem yet: different states
with positive or negative energy do not talk to each other and the evolution operator is perfectly unitary. 
A detailed discussion of this issue for the simplest higher-derivative system, the Pais-Uhlenbeck oscillator,
along with a critisism of somewhat confusive recent Bender and 
Mannheim papers \cite{BM} can be found in \cite{jaPU}.

  Ghosts usually still strike back for interacting nonlinear higher-derivative systems. The copious 
creation of positive energy and negative energy states brings about the {\it collapse} such that 
unitarity is violated, indeed. For this to happen, one does not need to have higher derivatives 
--- the simplest system
where this occurs is the quantum problem of 3D motion in an attractive potential $V(r) = -\alpha/r^2$ with large 
enough  $\alpha$. Falling on the center in this problem means breaking of unitarity
\cite{Popov}. Note that the associated {\it classical} problem is also sick: there are trajectories that fall 
on the center in finite time. Note also that the classical problem is sick for any positive $\alpha$, while 
the quantum one starts  having trouble only when $\alpha \geq \hbar^2/(8m)$. This 
is a rather typical universal situation: whenever quantum problem is sick, so is the classical one. 
The inverse is not true. Quantum fluctuations can sometimes cope with the singularity and prevent the system
from falling there. In other words, if we are interested with the state of health of some quantum system, we can
as well explore its classical brother. If the latter is OK, so is the former.

A natural nonlinear generalisation of the Pais-Uhlenbeck oscillator,
 \be
\lb{PUnonlin}
L \ =\ \frac 12(\ddot{q}^2 + \Omega^2 q^2)^2 - \frac \alpha 4 q^4 - \frac \beta 2 q^2 \dot{q}^2 \, ,
 \ee
was considered in \cite{benmal}. We have unravelled the presence of 
``islands of stability'' ---
in a certain range of the parameters $\alpha, \beta$ and for small 
enough initial fluctuations
$q(0), \dot{q}(0), \ddot{q}(0), q^{(3)}(0)$, the trajectories 
do not display any collapse, but oscillate 
near the perturbative vacuum $q=0$. However, when the deviations are large 
enough, the trajectories 
go astray and hit infinity. Thus, in spite of the presence of some benign 
trajectories, such model 
is not benign as a whole.

A completely benign higher-derivative supersymmetric quantum mechanical 
model was constructed in
\cite{Robert}. Its action has the form
  \be
\lb{actQM}
S \ =\ \int dtdx d\bar\theta d\theta \left[ \frac i2 \bar {\cal D} \Phi \frac {d}{dt}
 {\cal D} \Phi + V(\Phi) \right] \, ,
\ee
where
 \be
\lb{Phi}
\Phi = \ \phi + \theta \bar \psi + \psi \bar \theta + D \theta \bar\theta
 \ee
is a real (0+1)-dimensional superfield. The Lagrangian in \p{actQM} 
has an extra time derivative compared to  well-known Witten's SQM 
Lagrangian \cite{Witten}. As a result,
the former auxiliary field $D$ becomes dynamical.
The bosonic part of the component Lagrangian reads
 \be
\label{LB}
L_B \ =\ \dot{\phi} \dot{D} + D V'(\phi) \, .
 \ee
The canonical bosonic Hamiltonian involves now two pairs of
dynamic variables: $(\phi, p)$ and $(D, P)$ \ 
\footnote{One could, of course, get rid of the derivative $\dot{D}$ by adding a total derivative to the Lagrangian \p{LB}.
In this case $D$ becomes a Lagrange multiplier. But then the second derivative $\ddot{\phi}$ would appear 
in the Lagrangian such that the number of degrees of freedom stays the same.},
   \be
\label{HB}
H_B = pP - DV'(\phi) \, .
 \ee
It is not positive definite, and its spectrum does not have a bottom. Still, the 
dynamics of the system is completely benign. In fact, it involves {\it two} integrals of motion
(dimensionally reduced versions of the expressions \p{N} and \p{E} below), and is thus integrable.
Fot the simplest nontrivial superpotential,
 \be
\lb{VPhi}
 V(\Phi) = \ -\frac {\omega^2  \Phi^2}2 - \frac {\lambda \Phi^4}4 \, ,
 \ee
the solutions are expressed in terms of Jacobi elliptic functions (see equations \p{phit}, \p{param} below). 
Also the wave functions of the 
quantum states (with arbitrary high and arbitrary low energies) were found explicitly. The spectrum is 
continuous with eigenvalues lying in two intervals $] - \infty, -\omega] \cup [\omega, +\infty [$ plus the eigenvalue
$E=0$. The existence of zero energy eigenstates (infinitely many of them) annihilated by the action of the supercharges 
may be interpreted as the absence of spontaneous supersymmetry breaking. On the other hand, zero energy states 
are no longer {\it ground} states \dots 

Thus, benign {\it quantum-mechanical} systems with ghosts do exist. 
\footnote{It is not only \p{actQM}. An example of a stable 
nonlinear system with ghosts 
suggested back in 2003 \cite{Carroll} was  discussed in details 
in recent \cite{Kovner}.
Other examples were found in
\cite{Pavsic}.} However, no benign ghost-ridden {\it field theory} models were known up to now.
We will present an example of such a model below.

\section{A benign $2D$ higher-derivative model.} 
 The model is a straightforward two-dimensional generalization of \p{actQM}. We take the real superfield
\p{Phi} and assume now that its components depend not only on $t$, but also on $x$. 
The superfield action is
  \be
\lb{act2D}
S \ =\ \int dt dx d\bar\theta d\theta  \left[ -2i {\cal D} \Phi \partial_+ 
{\cal D} \Phi + V(\Phi) \right] \, ,
 \ee
where $\partial_\pm = (\partial_t \pm \partial_x)/2$ and 
 \be
\lb{DDbar}
{\cal D} = \frac {\partial}{\partial \theta} + i\theta \partial_-, \ \ \ \ \ \ \ \ \ \ 
\bar {\cal D} = \frac {\partial}{\partial \bar\theta} - i \bar\theta \partial_+
 \ee
are the $2D$ supersymmetric covariant derivatives.
\footnote{See \cite{West} for a good pedagogical description of the $2D$ superfield formalism.}
Note that the first term in \p{act2D} can as well be written as 
$-2i \bar {\cal D} \Phi \partial_- \bar {\cal D} \Phi$. The integrals over $ d\bar\theta d\theta$ of these two
expressions coincide. 

Expressing \p{act2D} in the components, we obtain
 \be
\lb{L2D}
{\cal L} = \partial_\mu \phi  \, \partial_\mu D \, + \, \partial_\mu \bar\psi  \, \partial_\mu \psi
+ DV'(\phi) \, + \, V''(\phi) \bar \psi \psi \, .
 \ee
Its dimensional reduction gives the Lagrangian studied in \cite{Robert}. 
The bosonic part of \p{L2D} is 
 \be
\lb{L2DB}
{\cal L}_B \ =\ \partial_\mu \phi  \, \partial_\mu D + DV'(\phi)\, .
 \ee
When it is benign, the whole system is benign.
\footnote{The latter is, however, not so relevant for us in this paper, where our primary goal is to 
construct {\it an} example of benign theory with ghosts. The Lagrangian \p{L2DB} 
is already such an example.}

Let us 
study the classical dynamics of \p{L2DB} with the superpotential \p{VPhi}. 
The equations of motion are
 \be
\lb{eqmot}
\Box \phi + \omega^2 \phi + \lambda \phi^3 &=& 0 \nn
\Box D + D(\omega^2 + 3\lambda \phi^2) &=& 0 \, .
 \ee
We see that $\phi(x,t)$ satisfies a nonlinear wave equation. The solutions to this equation cannot grow
--- the positive definite integral of motion
\be
\lb{N} 
N \ =\ \int dx \left[ \frac 12 \left( \dot{\phi}^2 + (\phi')^2 \right) + \frac {\omega^2 \phi^2}2
+ \frac {\lambda \phi^4}4 \right] 
 \ee
does not allow this. 
The equation for $D(x,t)$ is more tricky. $D(x,t)$ can grow with time, but we shall shortly see that 
this growth is at worst linear and does not lead to collapse.
Besides \p{N}, the system has also the usual energy integral of motion,
 \be
\lb{E}
E \ =\ \int dx \left[ \dot{\phi} \dot{D} + \phi' D' + D\phi(\omega^2 + \lambda \phi^2) \right]\,,
\ee
which can be both positive and negative.

    Two integrals of motion are not enough to make the field dynamics regular,
 and the latter exhibits chaotic features. We are in a position to solve
 the equations \p{eqmot} numerically. We played with different values of the parameters 
$\omega, \lambda$ and with different initial conditions and never found a collapse, but only, at worst, a linear
growth of $D(x,t)$ with time. A typical behaviour is displayed in Fig. 1 where the dispersion
$d(t) = \sqrt{\langle D^2 \rangle_x}$ is plotted as a function of time when we have chosen $\omega = \lambda = 1$ and the initial 
conditions
 \be
\lb{incon}
\phi(x,t=0) &=& Ce^{-x^2},   \ \ \ {\rm with} \ \ \ \ C=1,3,5 \nn
D(x,t=0)    &=& \cos{\pi x/L}\,,  \ \ \ \ \ \ \ \ \ \ L =10 \ \ {\rm being\ the\ length\ of\ the\ box}
 \ee

\begin{figure}[ht!]
     \begin{center}
        \subfigure[$C=1$]{%
            \label{hren1}
            \includegraphics[width=0.4\textwidth]{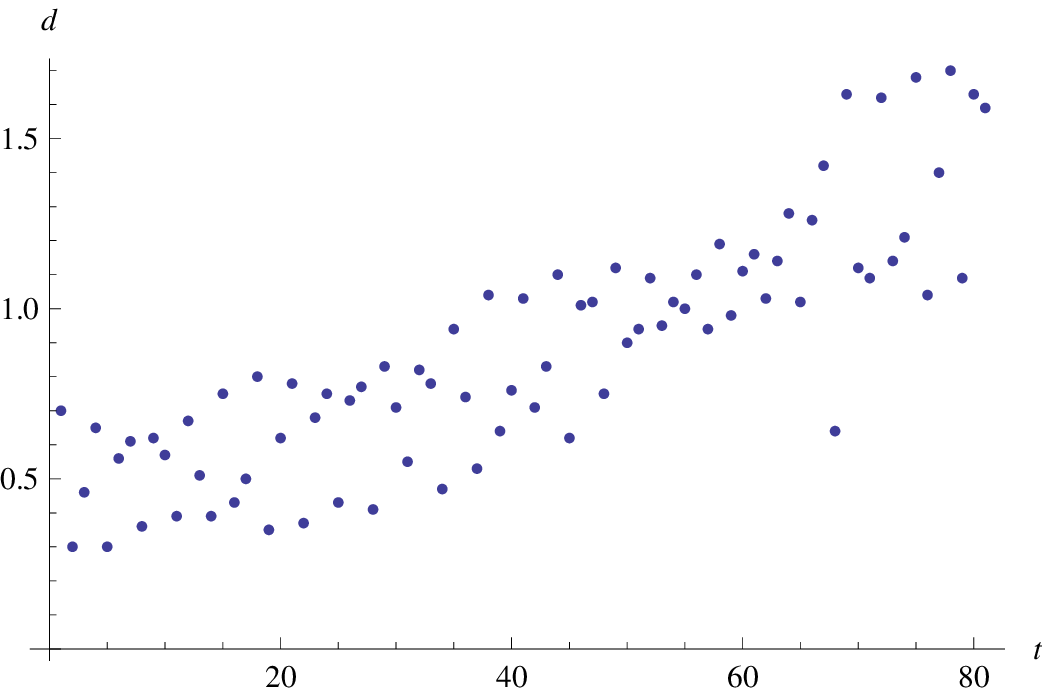}
        }%
        \subfigure[$C=3$]{%
           \label{hren2}
           \includegraphics[width=0.4\textwidth]{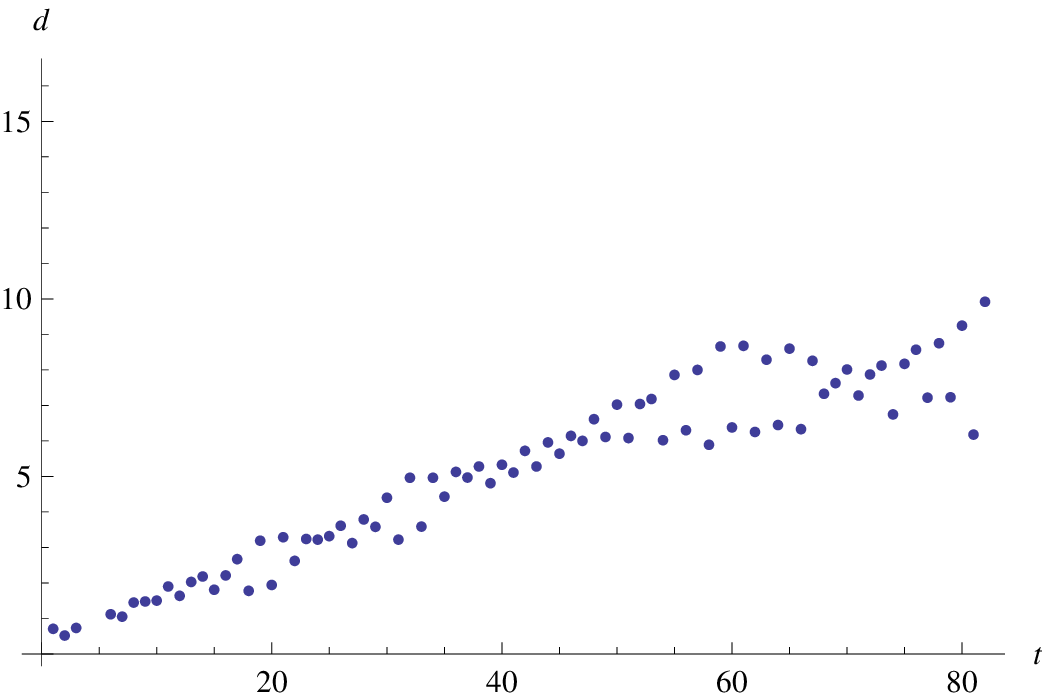}
        }
\subfigure[$C=5$]{%
           \label{hren3}
           \includegraphics[width=0.4\textwidth]{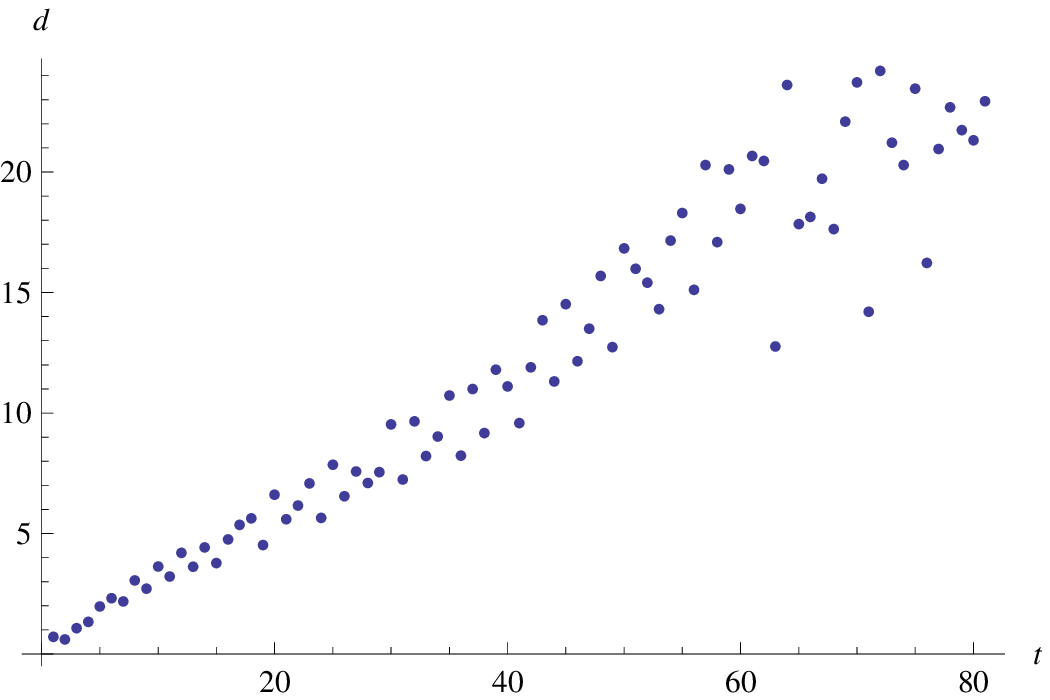}
        }
    \end{center}
    \caption{ Dispersion $d = \sqrt{\langle D^2 \rangle}$ as a function of time
with the initial conditions \p{incon} for different values of $C$.}
    \end{figure}

Due to chaoticity, $d$ is not a smooth function of $t$, 
but exhibits fluctuations. Still the linear growth trend is clearly seen.
 Note, however,  that this growth is in a considerable extent a finite size effect.
The initial conditions \p{incon} are chosen such that the energy density (the integrand in \p{E}) represents
a simple lump concentrated near zero.
But when time grows, this lump starts to travel to the left and to the right and soon reaches the boundary.
With larger values of $L$, this happens later, 
which affects the dynamics of $\phi(x,t), D(x,t)$, and $d(t)$. In particular, the latter
grows slower (see Fig.2)

\begin{figure}[ht!]
     \begin{center}

            \includegraphics[width=0.6\textwidth]{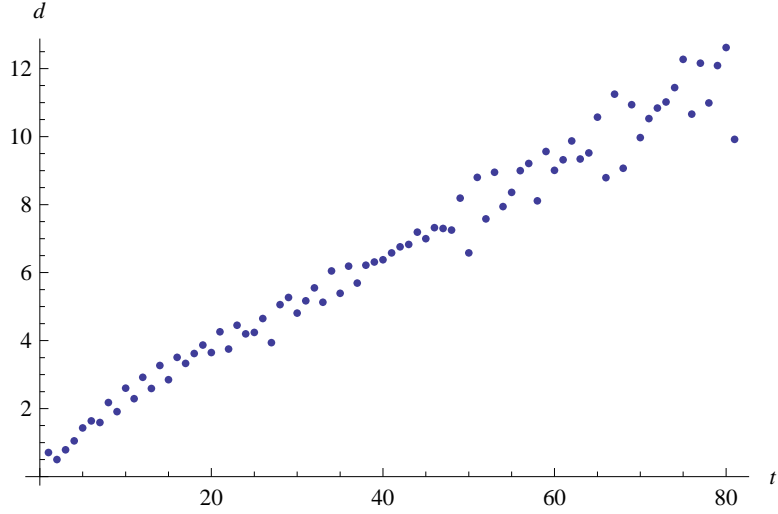}
        
    \end{center}
    \caption{ The same with $C=5$, $L=20$.}
    \end{figure}

\subsection{Possible implications for inflation.}

Another interesting choice for initial conditions is the {\it homogeneous} $\phi(x,t) \equiv \phi(t)$. The equation for 
$\phi(t)$ can in this case be solved analytically \cite{Robert}. It is an elliptic cosine function,
 \be
\lb{phit}
\phi(t) \ =\ \phi_0 \, {\rm cn}[\Omega t | m]
 \ee
with 
 \be
\lb{param}
\alpha = \frac {\omega^4}{\lambda N}, \ \ \Omega = [\lambda N(4+\alpha)]^{1/4}, \ \ \  m = \frac 12 \left[
1- \sqrt{\frac {\alpha}{4+\alpha}} \right]\, , \nn
\phi_0 \ =\ \left( \frac N \lambda \right)^{1/4} \sqrt{\sqrt{4+\alpha} - \sqrt{\alpha}} 
 \ee
where $N$ is the value of the integral of motion \p{N} and we used the {\it Mathematica} conventions.

With homogeneous $\phi(x,t)$, we can expand $D(x,t)$ in a Fourier series such that the second equation in \p{eqmot} splits
into independent equations for each Fourier component. The solutions to all these equations can also be expressed 
semianalytically via certain integrals \cite{Robert}, or else found numerically.
One way or another one observes
 that all nonzero Fourier modes of $D(x,t)$ stay bounded, and only the zero mode exhibits a linear growth
(see Fig. 3). As a result, the function $D(x,t)$ becomes more and more homogeneous.

\begin{figure}[ht!]
     \begin{center}
        \subfigure[$k=0$]{%
            \label{hren11}
            \includegraphics[width=0.4\textwidth]{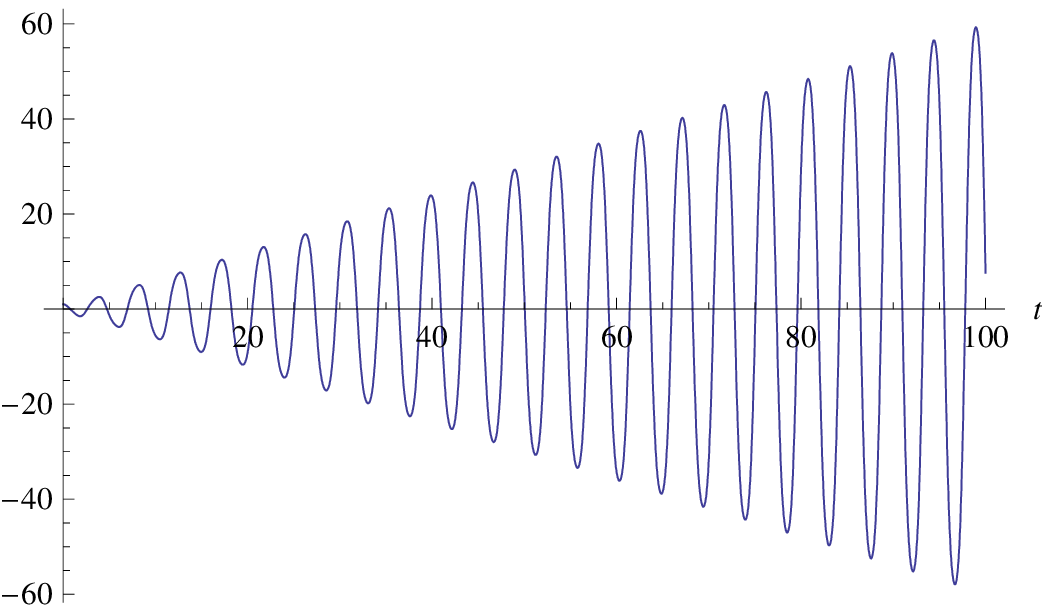}
        }%
        \subfigure[$k=1$]{%
           \label{hren22}
           \includegraphics[width=0.4\textwidth]{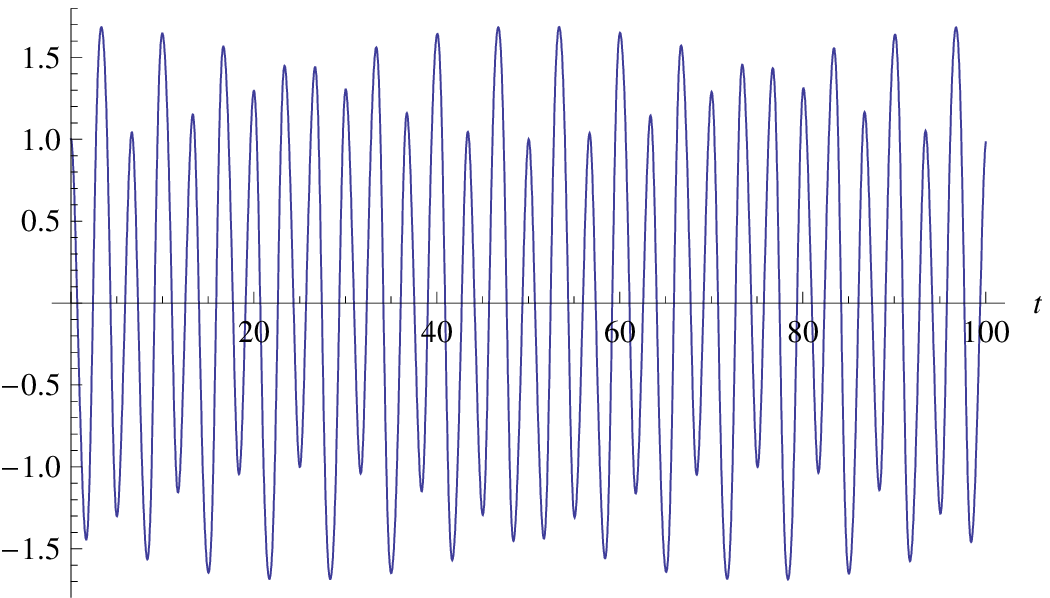}
        }

    \end{center}
    \caption{ (a) The growth of the zero Fourrier mode and (b) the boundness of the mode with nonzero $k$ for homogeneous $\phi(x,t) \equiv \phi(t)$. 
The parameters   $\omega=\lambda = N = 1$ are chosen.}
    \end{figure}

This phenomenon may be relevant for the  eventual solution of  the long-standing problem of {\it inflation initial conditions}. 
For the inflation to take place,  the scalar field which drives it should be homogeneous. 
At present, such homogeneity is mostly just postulated. But maybe it appears as a result of the evolution of a 
higher-derivative system at the stage {\it preceding} inflation ? 

This speculative idea expressed first in \cite{benmal} seems to us attractive. One should also mention, however, that,
as far as the toy model considered in this paper is concerned, the profiles for $\phi(x,t)$ and $D(x,t)$ do not become
homogeneous with arbitrarily chosen initial conditions. It is only when $\phi(x,t)$ was  homogeneous at the initial moment that 
$D(x,t)$ becomes homogeneous as time passes. Certainly, more studies in this direction are necessary...

\section{Conclusions and outlook.}
In this paper, we described the first example of  field theory which stays unitary in spite of the presence of ghosts.
One can guess that there are other such models, also with number of spatial dimensions higher than one and probably also higher
than three. In Ref. \cite{TOE}, we suggested
that one of such higher-dimensional higher-derivative models can play the role of the Holy Grail Theory of Everything such
that our (3+1)-dimensional Universe represents a classical brane-like classical solution of this 
theory embedded in a flat higher-dimensional bulk. My hopes at that time were associated
with a beautiful superconformal (at the classical level) 
6-dimensional renormalisable SYM theory\cite{ISZ}. It has a nontrivial and nicely looking 
Lagrangian and is asympthotically free.
 Unfortunately,  beauty does not always mean efficiency.
 This theory involves collapsing classical trajectories
\footnote{ It is especially clearly seen in the sector of the 
(former auxiliary) fields $D^a_{i=1,2,3}$ of canonical dimension 2 which enter the Lagrangian 
(for the $SU(2)$ gauge group) as
 \be
\lb{LDabc}
{\cal L}_D \propto \frac 12 (\dot D^a_i)^2 + \frac 16 \epsilon^{abc} \epsilon_{ijk} D^a_i D^b_j D^c_k\, .
 \ee
Substituting the ansatz $D^a_i = D\delta^a_i$ in the equations of motion, we obtain the equation
$\ddot D - D^2 = 0$, whose solutions are obviously singular.}
and 
is not {\it benign} in the sense outlined above. 
One should search for something else...


\begin{thebibliography}{96}

\bibitem{Isham} C.J. Isham, {\it Canonical quantum gravity and the problem of time}, 
gr-qc/9210011, published in the Proceedings of GIFT Int. Seminar on Theor. Physics, Salamanca, 15-27 June, 1992.

\bibitem{dragon} A.V. Smilga, {\it Quantum gravity as Escher's dragon}, Phys. Atom. Nucl. {\bf 66} (2003) 2092, hep-th/0212033.

\bibitem{Bern} Z. Bern et al, {\it Amplitudes and ultraviolet behavior of ${\cal N} = 8$ supergravity},
Fortsch. Phys. {\bf 59} (2011) 561, arXiv:1103.1848[hep-th].

\bibitem{Bossard} G. Bossard, P.S. Howe, and K.S. Stelle, {\it The Ultra-violet question 
in maximally supersymmetric field theories}, Gen. Rel. Grav. {\bf 41} (2009) 919, arXiv:0941.4661[hep-th].

\bibitem{Beisert} N. Beisert et al, {\it E7(7) constraints on counterterms in ${\cal N} = 8$ supergravity}, 
Phys. Lett. {\bf B694} (2010) 265, arXiv:1009.1643[hep-th].

\bibitem{Cederwall} M. Cederwall and A. Karlsson, {\it Loop 
amplitudes in maximal supergravity with manifest supersymmetry}, 
JHEP {\bf 1303} (2013) 114, arXiv:1212.5175[hep-th].


\bibitem{Stelle} K.S. Stelle,  {\it Renormalization of higher-derivative quantum gravity}, 
Phys. Rev. {\bf D16} (1977) 953.

\bibitem{Fradkin} E.S. Fradkin, A.A. Tseytlin, {\it Renormalizable 
asymptotically free quantum theory of gravity}, Nucl. Phys. 
{\bf B201} (1982) 469.

\bibitem{Sakharov} A.D. Sakharov, {\it Vacuum quantum fluctuations in curved 
space and the theory of gravitation}, Sov. Phys. Dokl. {\bf 12}
(1968) 1040.

\bibitem{Zvenigorod} A.V. Smilga, {\it Spontaneous generation of the Newton 
constant in the renormalizable gravity theory}, Preprint ITEP-63-1982,
published in the Proceedings of the conference on {\sl Group theoretical methods in physics} ( Zvenigorod, 1982), vol.2, p.73.

\bibitem{Adler} S.L. Adler, {\it Einstein gravity as a symmetry breaking effect 
in quantum field theory}, Rev. Mod. Phys. {\bf 54} (1982) 729.

\bibitem{Nieder} M. Niedermaier, {\it Gravitational fixed points from perturbation theory}, Phys. Rev. Lett. {\bf 103}
(2009) 101303. 

\bibitem{Mann} P.D. Mannheim, {\it Making the case for conformal gravity}, Found. Phys. {\bf 42} (2012) 388, 
arXiv:1101.2186[hep-th].

\bibitem{Anca} J. Kluso\v{n}, M. Oksanen, and A. Tureanu, {\it Hamiltonian analysis 
of curvature-squared gravity with or without conformal invariance}, arXiv:1311.4141 [hep-th].

\bibitem{PU}   A. Pais  and G.E. Uhlenbeck, {\it On field theories with nonlocalized action}, 
 Phys. Rev  {\bf 79} (1950) 145.

\bibitem{BM} C.M. Bender  and P.D. Mannheim, {\it No ghost theorem 
for the fourth-order derivative Pais--Uhlenbeck oscillator model},
Phys. Rev. Lett.  {\bf 100} (2008) 110402, arXiv:0706.0207[hep-th]; {\it Exactly solvable PT-symmetric Hamiltonian having no 
Hermitian counterpart}, Phys. Rev. {\bf D78} (2008) 025022, arXiv:0804.4190[hep-th].

\bibitem{jaPU} A.V. Smilga, {\it Comments on the dynamics of 
the Pais-Uhlenbeck oscillator}, SIGMA {\bf 5} (2009) 017, arXiv:0808.0139[quant-ph].

\bibitem{Popov} A.M. Perelomov and V.S. Popov, {\it Collapse onto scattering center in
quantum mechanics}, Teor. Mat. Fiz. {\bf 4} (1970) 48.


\bibitem{benmal} A.V. Smilga, {\it Benign versus malicious ghosts in higher-derivative theories},
Nucl. Phys. {\bf B706} (2005) 598,  hep-th/0407231 .

\bibitem{Robert} D. Robert and A.V. Smilga, {\it Supersymmetry versus ghosts}, J. Math. Phys.
{\bf 49} (2008) 042104, math-ph/0611023.

\bibitem{Witten} E. Witten, 
{\it Dynamical breaking of supersymmetry}, Nucl. Phys. {\bf B188} (1981) 513.

\bibitem{Carroll} S.M. Carroll, M. Hoffman, and M. Trodden, {\it Can the dark energy 
equation-of-state parameter $w$ be less that -1?}, Phys. Rev. {\bf D68} (2003) 023509, 
astro-ph/0301273.

\bibitem{Kovner}   I.B. Ilhan and A. Kovner, {\it Some comments on ghosts and unitarity: 
the Pais-Uhlenbeck oscillator revisited},  Phys. Rev.  {\bf D88} (2013) 044045, arXiv:1301.4879[hep-th].

\bibitem{Pavsic} M. Pavsic, {\it Stable self-interacting Pais-Uhlenbeck oscillator},
Mod. Phys. Lett. A 28 (2013) 1350165, arXiv:1302.5257[gr-qc].

\bibitem{West} P.C. West, {\it Introduction to supersymmetry and supergravity}, World Scientific, 
Singapore, 1986.

\bibitem{TOE} A.V. Smilga {\it $6D$ superconformal theory as the Theory of Everything}, Czech J. Phys. {\bf 56} (2006) C443,
arXiv:hep-th/0509022. 

\bibitem{ISZ} E.A. Ivanov, A.V. Smilga, B.M. Zupnik, {\it Renormalizable supersymmetric gauge theory in six dimensions}, 
Nucl. Phys. {\bf B726} (2005) 131,  hep-th/0505082.

\end{thebibliography}
\end{document}